\documentclass[
 preprint,
 superscriptaddress,
 amsmath,amssymb,
 aps,
 prl,
 floatfix
]{revtex4-1}

\usepackage{graphicx}% Include figure files
\usepackage{dcolumn}% Align table columns on decimal point
\usepackage{bm}% bold math
\usepackage[colorlinks,citecolor=blue]{hyperref}

\begin{document}

\title{Hot Polarons with Trapped Excitons and Octahedra-Twist Phonons in $\mathrm{CH_{3}NH_{3}PbBr_{3}}$ Hybrid Perovskite Nanowires}

\author{Feilong Song}
\author{Chenjiang Qian}
\affiliation{Beijing National Laboratory for Condensed Matter Physics, Institute of Physics, Chinese Academy of Sciences, Beijing 100190, China}
\affiliation{CAS Center for Excellence in Topological Quantum Computation and School of Physical Sciences, University of Chinese Academy of Sciences, Beijing 100049, China}
\author{Yunuan Wang}
\affiliation{Beijing National Laboratory for Condensed Matter Physics, Institute of Physics, Chinese Academy of Sciences, Beijing 100190, China}
\affiliation{Key Laboratory of Luminescence and Optical Information, Ministry of Education, Beijing Jiaotong University, Beijing 100044, China}
\author{Feng Zhang}
\affiliation{Beijing Key Laboratory of Nanophotonics and Ultrafine Optoelectronic Systems, School of Materials Science $\&$
Engineering, Beijing Institute of Technology, Beijing, 100081, China}

\author{Kai Peng}
\author{Shiyao Wu}
\author{Xin Xie}
\author{Jingnan Yang}
\author{Sibai Sun}
\author{Yang Yu}
\author{Jianchen Dang}
\author{Shan Xiao}
\author{Longlong Yang}
\affiliation{Beijing National Laboratory for Condensed Matter Physics, Institute of Physics, Chinese Academy of Sciences, Beijing 100190, China}
\affiliation{CAS Center for Excellence in Topological Quantum Computation and School of Physical Sciences, University of Chinese Academy of Sciences, Beijing 100049, China}
\author{Kuijuan Jin}
\affiliation{Beijing National Laboratory for Condensed Matter Physics, Institute of Physics, Chinese Academy of Sciences, Beijing 100190, China}
\affiliation{CAS Center for Excellence in Topological Quantum Computation and School of Physical Sciences, University of Chinese Academy of Sciences, Beijing 100049, China}
\affiliation{Songshan Lake Materials Laboratory, Dongguan, Guangdong 523808, China}
\author{Haizheng Zhong}

\affiliation{Beijing Key Laboratory of Nanophotonics and Ultrafine Optoelectronic Systems, School of Materials Science $\&$
Engineering, Beijing Institute of Technology, Beijing, 100081, China}

\author{Xiulai Xu}
\email{xlxu@iphy.ac.cn}
\affiliation{Beijing National Laboratory for Condensed Matter Physics, Institute of Physics, Chinese Academy of Sciences, Beijing 100190, China}
\affiliation{CAS Center for Excellence in Topological Quantum Computation and School of Physical Sciences, University of Chinese Academy of Sciences, Beijing 100049, China}
\affiliation{Songshan Lake Materials Laboratory, Dongguan, Guangdong 523808, China}

\begin{abstract}

Hybrid Perovskites have shown a great potential for applications in photovoltaics and light-emitting devices with high efficiency. Interaction between defect-induced trapped excitons and phonons plays an important role in understanding the emerging phenomena for such an excellent figure-of-merit. Here we demonstrate hot polarons with narrow linewidth in $\mathrm{CH_{3}NH_{3}PbBr_{3}}$ nanowires, which originate from the interaction between trapped excitons and octahedra-twist phonons. The observation of hot polarons in photoluminescence without gain methods indicates the large interaction strength between excitons and phonons. The multiple hot polarons are further confirmed by magneto-optical spectra with a Zeeman splitting of the trapped excitons and a phonon energy increase with diamagnetic effect. Furthermore, the phonons participating in the interaction are demonstrated to be the octahedra-twist vibrations which are transverse optical phonons, while the interaction between trapped excitons and longitudinal optical phonons is weak. Our work demonstrates that trapped excitons in perovskites prefer to interact with transverse rather than longitudinal optical phonons. Since bulk materials usually interact with longitudinal optical phonons, this result provides a physical explanation of the high tolerance of defects in perovskites.

%\pacs{42.50.Pq,78.67.Pt,78.67.Hc}
\end{abstract}

\maketitle

\section{\label{sec1} Introduction}

Hybrid organic-inorganic perovskites (HOIPs) are a potential material with attractive optoelectronic properties such as high luminescence efficiency, high power converting efficiency and long carrier diffusion lengths \cite{doi:10.1002/adma.201502294,Brenner2016,Li2017}. Thus, HOIPs are widely applied in high efficient solar cells, high quantum yield LEDs, and ultra low threshold lasers \cite{Heo2013,Jeon2014,Xing2014,C4CS00458B,doi:10.1021/acs.jpclett.9b00658}. These advantages are generally ascribed to the high tolerance of defects \cite{doi:10.1021/acs.nanolett.6b02688,doi:10.1021/acsenergylett.7b00547}. However, the defects or defect-induced trapped localized excitons (LXs) in HOIPs have only been observed with an ensemble behavior from the broad photoluminescence (PL) spectrum \cite{PhysRevB.97.134412,doi:10.1021/acs.jpclett.7b02979}. As a result, many questions about the physical processes with defects and LXs remain open.
%As a result, physics behind the defects and trapped excitons are not very clear yet.

Besides the defects, the interaction between excitons and phonons is also important for optoelectronic properties such as carrier mobility \cite{doi:10.1146/annurev-physchem-040215-112222,RevModPhys.89.015003,doi:10.1021/acsnano.7b05033}. The exciton-phonon polaron has been intensively investigated in various materials including HOIPs. While these investigations mainly focus on the longitudinal optical (LO) phonon due to the stronger interaction to excitons, e.g., interactions between carriers and LO phonons are usually thought to prevent defect scatterings and thereby result in the high tolerance of defects \cite{Zhu1409,doi:10.1021/acs.jpclett.5b02462,Miyatae1701217}. In contrast, HOIPs have unique orthorhombic structures in which a Pb atom is caged by octahedral halogen atoms. The cage has various octahedra-twist vibrations as transverse optical (TO) phonons with smaller energy compared to LO phonons \cite{doi:10.1021/acs.jpclett.7b02979,C6CP03474H}. Interaction between TO phonons and excitons in HOIPs has rarely been investigated so far, which could provide a deep understanding of the emerging optoelectronic properties for perovskites.

Here we demonstrate hot polarons originating from the interaction between LXs and TO phonons in HOIP $\mathrm{CH_{3}NH_{3}PbBr_{3}}$ nanowires. At low temperature, a zero-phonon LX with narrow linewidth is observed, along with equidistant hot polarons at high energy side with increasing excitation power.  The phonon has a small energy corresponding to octahedra-twist vibrations \cite{doi:10.1021/acs.nanolett.7b00064,doi:10.1021/acs.jpclett.7b02979}. Distinctive magneto properties have been observed with Zeeman splitting and diamagnetic effects from the LX and hot polaron replicas. In contrast, the interaction between LXs and LO phonons is observed contrastly weak. While free excitons (FX) or carriers in bulk materials are usually coupled to LO phonons \cite{Zhu1409,doi:10.1021/acs.jpclett.5b02462,Miyatae1701217}. The different vibration (phonon) modes of FXs and LXs reduce the phonon-assisted interaction between defects and bulk materials, providing a further explanation to the high tolerance of defects in HOIPs.

\section{Results and discussion}

The HOIP nanowires were colloidally synthesized by a ligand-assisted reprecipitation (LARP) method \cite{doi:10.1021/acsnano.5b01154,doi:10.1002/cnma.201700034}, namely, adding a solution of $\mathrm{CH_{3}NH_{3}PbBr_{3}}$ in solvent (N,N-dimethylformamide, DMF) drop-wise into a solvent (toluene) to induce the crystallization process. In this method, both nucleation and growth process simultaneously. Thus, the size and shape of HOIPs are hard to control. In contrast, the HOIP nanowires used in our work were grown by a "reverse" LARP method, changing the sequence of the solvent mixing by pouring toluene into the perovskite precursor solution in DMF. In this way, the nucleation and growth process separately. Thus, the size and shape of HOIPs are easy to control, resulting in the high quality of HOIP nanowires. The optical properties were measured with a conventional confocal micro-PL system. The device was mounted on a three-dimensional nano positioner and cooled to 4.2 K by exchanging helium gas with a helium bath. A HeCd laser with emission wavelength at 442 nm was used to excite HOIP nanowires. The PL spectra were collected by an area array CCD detectors with the resolution of 0.2 meV in the spectra range around 550 nm. A magnetic field was applied by a superconducting magnet with the maximum value of 9 T.

\begin{figure}
\centering
\includegraphics[width=\linewidth]{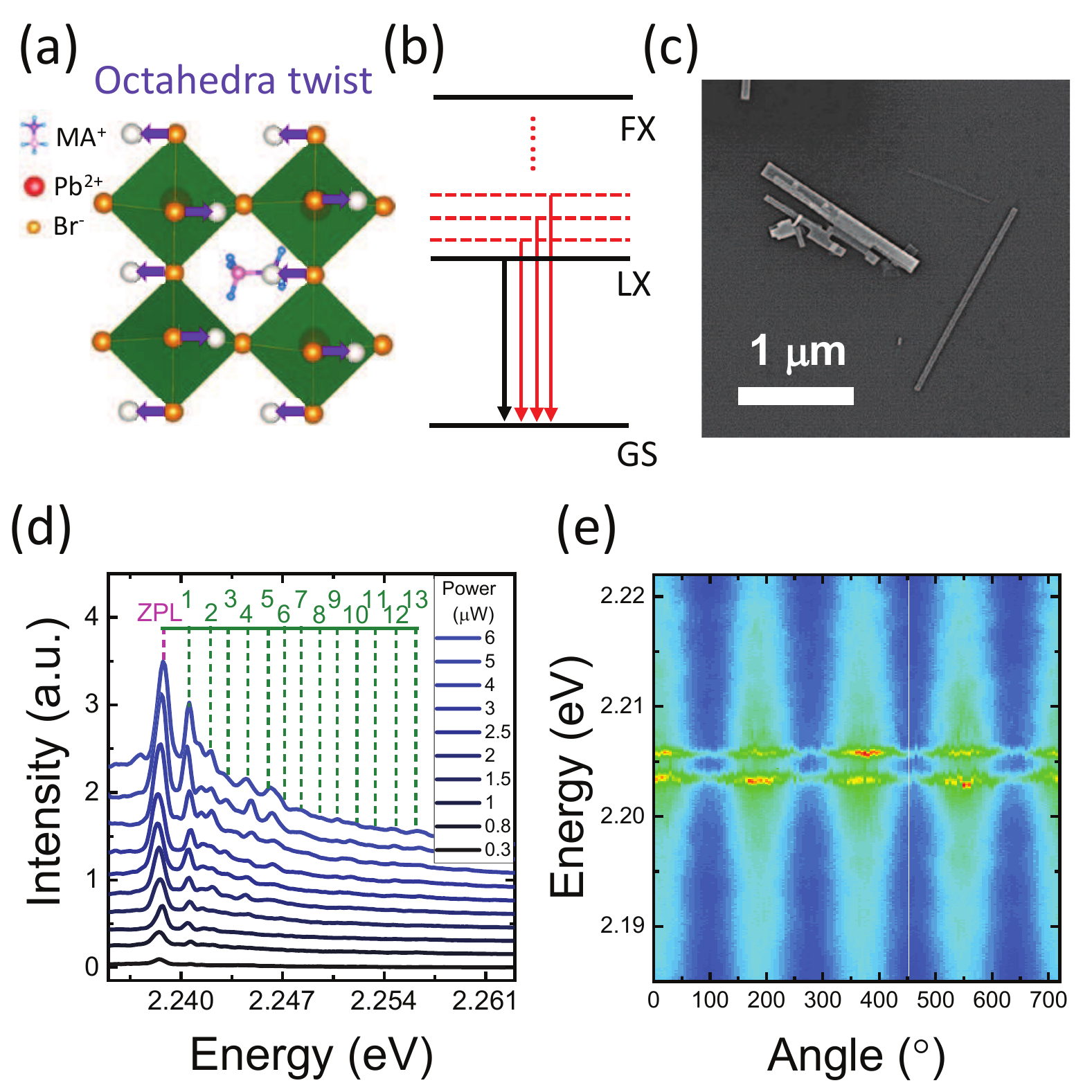}
\caption{\label{f1} (a) Orthorhombic crystal structure of MAPbBr$_{3}$ at low temperature ($T<140\ \mathrm{K}$). Yellow dots refer to Br atoms and green areas refer to octahedra cages. Purple arrows and gray dots refer to the cage twist vibration which is a TO phonon. (b) Schematic diagrams of energy levels. GS is ground state, FX and LX refer to free excitons and localized excitons respectively. (c) a SEM image of an ensemble of MAPbBr$_{3}$ nanowires. (d) Power-dependent PL spectra at 4.2 K. Red and green dished lines show the energy positions of the ZPL and hot polarons respectively. (e) Polarization-dependent spectra of one LX state and one polaron state, which are linearly polarized.}
\end{figure}

The structure of HOIPs under low temperature is the orthorhombic phase, as shown in Fig.~\ref{f1}(a). The $\mathrm{Pb}$ atoms are in the center (red dots), the $\mathrm{Br}$ atom cage is around the $\mathrm{Pb}$ atom (yellow dots and green cages) and the $\mathrm{CH_{3}NH_{3}}$ ($\mathrm{MA}$) fills the voids \cite{doi:10.1021/acs.jpclett.5b02462}. Fig.~\ref{f1}(c) shows a scanning electron microscope (SEM) image of the $\mathrm{MAPbBr_3}$ nanowires. Typically the nanowires have a rectangular cross section with a length of a few microns. Figure~\ref{f1}(d) shows excitation-power-dependent PL spectra for one nanowire. Under a low excitation power, only zero-phonon line (ZPL) is observed and the linewidth is $0.8\ \mathrm{meV}$. As the excitation power increases, more peaks (up to 13) at the high energy side with equal energy space appear as hot polarons, which are schematically sketched in Fig.~\ref{f1}(b). The emission of hot polarons is usually weak and has only been observed with extra gain methods such as nanocavity enhancement \cite{Cho2013}. Contrastly, the direct observation of hot polarons indicates the large interaction strength between excitons and phonons \cite{doi:10.1021/acs.nanolett.5b00109,doi:10.1021/jacs.6b08175,doi:10.1021/acsenergylett.7b00862,doi:10.1021/acs.jpclett.6b02046,doi:10.1002/aenm.201602174,C8EE03369B}. Two other possible explanations for emission peaks at high energy side are Rydberg exciton states \cite{Kazimierczuk2014} or Fabry-Perot (FP) cavity modes \cite{Zhu2015}, which can be excluded easily. For Rydberg exciton states, the excited states $E_n= E_g-{E_b}/{n^{2}}$ are not equidistant. Although FP modes are equidistant, the length of the nanowire $L=1\ \mathrm{\mu m}$ corresponds to the free spectral range $\lambda^{2}/2nL=59\ \mathrm{nm}$ with a center wavelength $\lambda =550\ \mathrm{nm}$ and the refraction index n of 2.55 for MAPbBr$_3$. This value around 200 meV is much more than the equal energy space around $1.7\ \mathrm{meV}$ here.

Figure~\ref{f1}(d) clearly shows the linear relevance between the hot polaron energy and the phonon number. The phonon energy is fitted with a value around $1.7\ \mathrm{meV}$, namely, the wave number (WN) of $14\ \mathrm{cm^{-1}}$. In HOIPs, the LO phonon usually has a large energy over $50\ \mathrm{cm^{-1}}$. The phonon energy of $14\ \mathrm{cm^{-1}}$ here corresponds to the octahedra-twist vibration of the cage, which is the vibration mode with the smallest optical phonon energy in HOIPs \cite{doi:10.1021/acs.nanolett.7b00064,doi:10.1021/acs.jpclett.7b02979}. The octahedra-twist vibration is a TO phonon as schematized by purple arrows in Fig.~\ref{f1}(a). Generally, TO phonons are hard to observe in bulk materials due to the phonon energy smaller than the exciton linewidth \cite{doi:10.1021/acs.jpclett.7b02979}. Thus, the narrow linewidth of ZPL here could be an important reason for the observation of hot polarons with TO phonons. As the linewidth of $0.8~\mathrm{meV}$ is much smaller than that of excitons in bulk materials, the ZPL cannot originate from self-trapping, because self-trapped free exciton usually has a broader linewidth. Therefore, the ZPL can be ascribed to the defect-induced LX \cite{PhysRevB.97.134412,doi:10.1021/acs.jpclett.7b02979}. Meanwhile, LXs are partially polarized due to the anisotropy of the nanowire geometry \cite{doi:10.1021/nl9017012}, as shown in Fig.~\ref{f1}(e).

\begin{figure}
\centering
\includegraphics[width=\linewidth]{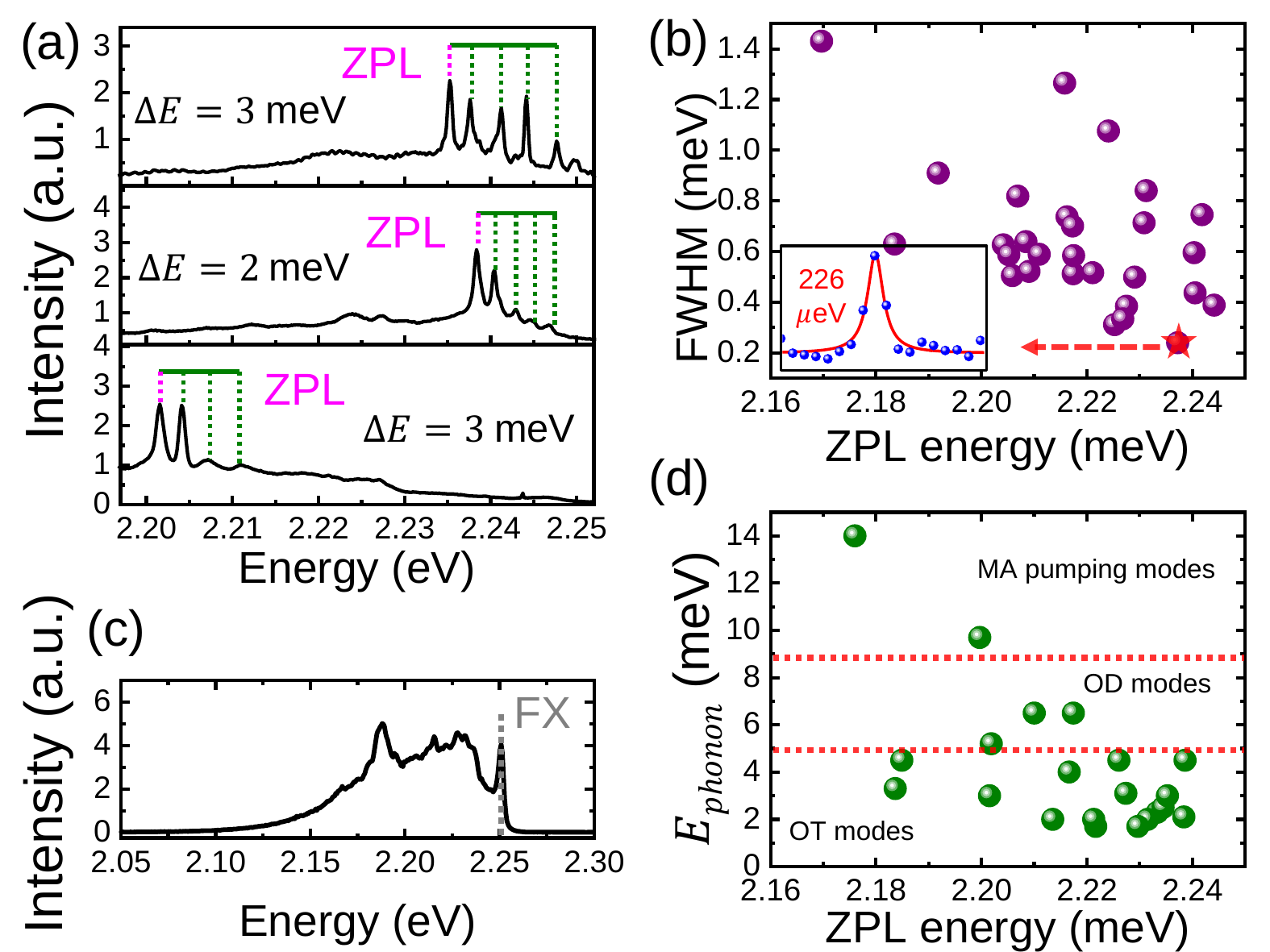}
\caption{\label{f2} (a) PL spectra of different nanowires. ZPL from the LX and hot polarons are marked. (b) Statistics of the linewidth. Inset shows a narrowest peak with a full width at half maximum (FWHM) of $226\ \mathrm{\mu eV}$. (c) Typical PL spectrum with FX energy at $2.25\ \mathrm{meV}$. The broad peak originates from the recombinations of the defect states. (d) Statistics of the phonon energy. Red dashed lines separate different vibration modes. Octahedra twist (OT) is the TO phonon mode. Octahedra distorting (OD) is TO/LO phonon modes. MA pumping modes are LO phonons. Gray dashed lines in (b) and (d) refer to FX position.}
\end{figure}

To confirm the origination of the ZPLs, PL spectra at various positions have been collected, as shown in Fig.~\ref{f2}(a) for three more locations from different wires. It can be seen that the ZPL energy varies within a wide range from 2.15 to 2.24 $\mathrm{eV}$. The observed ZPL energy statistics are shown in Fig.~\ref{f2}(b). This variation can not be ascribed to the quantum confinement of excitons in nanowires. The cross spatial length of nanowires is around 100 nm and much larger than the exciton Bohr radii of around 5 nm, thus, the quantum confinement effect is very weak \cite{RevModPhys.82.427,PhysRevB.94.125139}. Additionally, the ZPL energy is mucher smaller than that of FXs in $\mathrm{MAPbBr_{3}}$ (around $2.25\ \mathrm{eV}$ at 4K) as shown in Fig.~\ref{f2}(c) \cite{doi:10.1021/acs.jpclett.7b02979,doi:10.1021/acsnano.6b02734}. The broad peak in Fig.~\ref{f2}(c) originates from the recombinations defect states \cite{PhysRevB.97.134412,doi:10.1021/acs.jpclett.7b02979}. Therefore, the ZPLs can be confirmed as defect-induced LXs. Figure \ref{f2}(b) shows the statistics of the linewidth of the LX. The linewidth decreases with the ZPL energy might be explained by the defect size related lattice residual damage \cite{Chen2016}. Figure \ref{f2}(d) shows the statistics of the phonon energy, which also generally decreases with the ZPL energy. Combining Fig.~\ref{f2}(b) and (d), it can be clearly observed that the coupled phonon energy decreases with the ZPL linewidth, which means that LXs with narrower linewidth prefer to interact with TO phonons. This result agrees well with previous works that excitons with broad linewdith are observed to interact with LO phonons. The narrowest linewidth (inset in Fig.~\ref{f2}(b)) measured is about $226\ \mathrm{\mu eV}$, which is close to the resolution of our spectrometer. This is also the smallest value reported in HIOPs up to now \cite{PhysRevB.97.134412,doi:10.1021/acs.jpclett.7b02979}.

\begin{figure}
\centering
\includegraphics[scale=0.7]{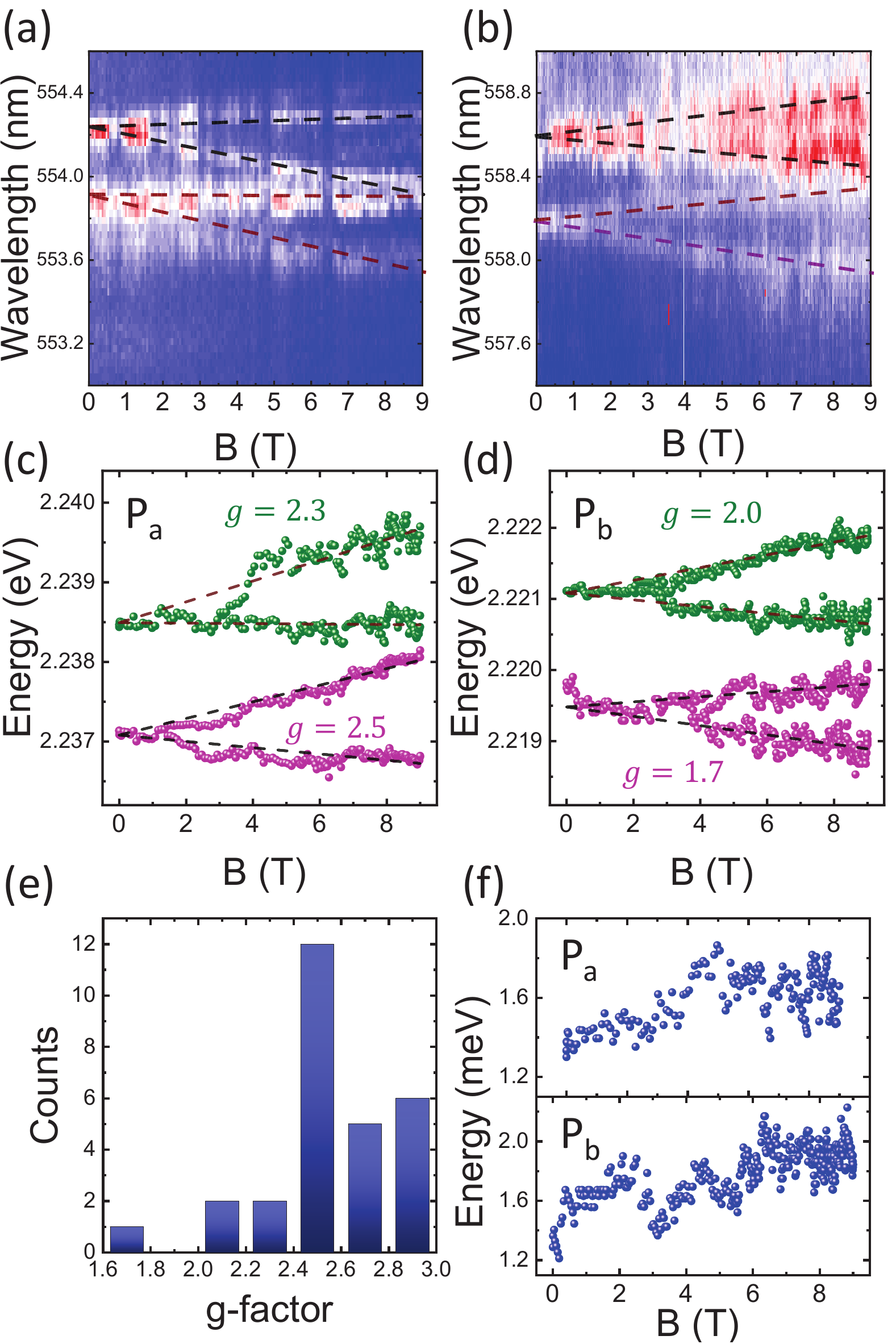}
\caption{\label{f3} (a)(b) Magneto-PL map at (a) position P$_{a}$ and (b) position P$_{b}$. Black dashed lines refer to the ZPL and wine dashed lines refer to the polaron. (c)(d) The extracted peak energies and the g-factors at position Pa (c) and position Pb (d). (e) Statistics of g factor of different HOIP nanowires. (f) The phonon diamagnetic effect at two positions. Phonon energies are calculated by $E_{phonon}=\overline{E}_{polaron}-\overline{E}_{ZPL}$, where $\overline{E}_{polaron}$ ($\overline{E}_{ZPL}$) is the average energy of the split two peaks of polaron (exciton) as shown in (c) and (d).}
\end{figure}

Magneto-optical properties are also very important for semiconductors, paving ways to investigate and control excitonic properties such as emission energies, polarizations and wavefunction distributions \cite{Cao2016,PhysRevLett.122.087401}. The PL spectra in a magnetic field of HOIP nanowires were collected with Faraday configuration (magnetic field parallel to the laser direction). Figure~\ref{f3}(a) and (b) show contour plots of PL spectra as function of applied magnetic field at two different positions P$_{a}$ and P$_{b}$. The Zeeman splitting of ZPL (black dashed lines) and the first polaron (wine dashed lines) can be clearly observed. The energies of each peak are extracted in Fig.~\ref{f3}(c) and (d) respectively. The g-factor of the polaron is close to the ZPL for both positions, which agree with to previous results of phonons \cite{Skolnick_1983}. The g-factors for different positions are shown in shown in Fig.~\ref{f3}(e). It can be seen that g-factor variates around 2, due to the zero-dimensional feature of LXs which is similar to quantum dots \cite{doi:10.1021/acs.nanolett.7b00064}. Additionally, the diamagnetic effects of phonon at two positions are extracted as shown in Fig.~\ref{f3}(f). The phonon energy increases with a value of around $0.5\ \mathrm{meV}$ at $B=9\ T$. The small diamagnetic is due to that the wave vector of the phonon is modified to offset the electron momentum with the vector potential to maintain momentum conservation \cite{Zhang_2014}. The diamagnetic within the magnetic filed range (9 T) here is relatively small, which needs further investigations in the future. Overall, the magneto-optical properties of polarons show the effects from both phonons and LXs, which results in a unique magnetic field control of excitonic system in HOIPs.

\begin{figure}
\centering
\includegraphics[width=\linewidth]{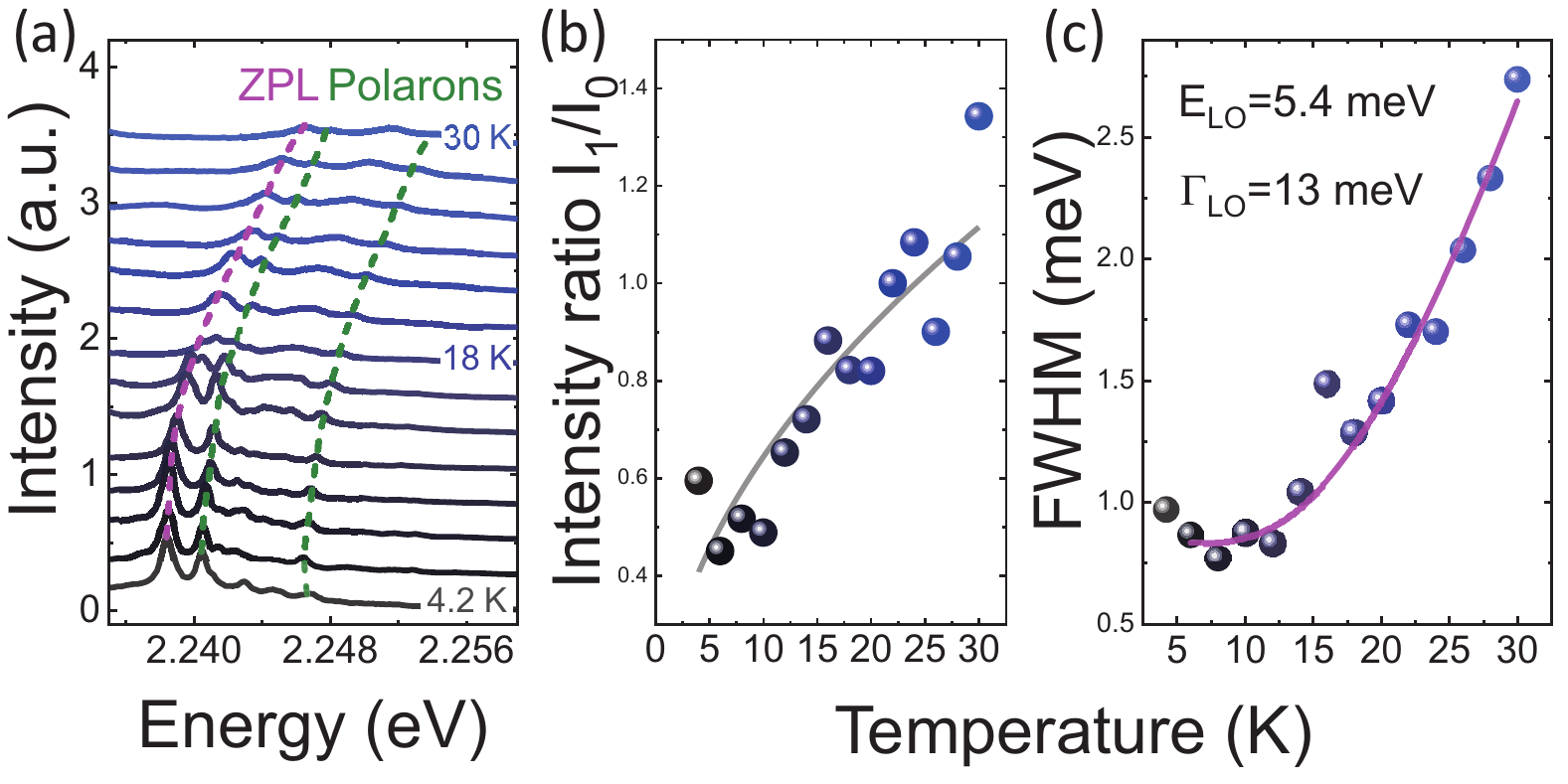}
\caption{\label{f4} (a) Temperature-dependent PL spectra from 4 K to 30 K. Large blue shift and intensity decreases can be observed with increasing temperature. (b) The intensity ratio between the first hot polaron (I$_1$) and the ZPL (I$_0$) as a function of temperature. Gray line is the square root $\sqrt{T}$ fitting. (c) Linewidth variation of the ZPL (symbols) and the fitted results. The interaction strength $\Gamma_{LO}$ is quite small, resulting in the unobserved LO phonon in the spectra.}
\end{figure}

Normally in most semiconductor materials, the exciton is usually coupled to LO phonon, due to that LO phonon could introduce relative displacement between positively charged atoms and negatively charged atoms \cite{Yu2010Fundamentals,PhysRevB.38.13377,Wright2016}. However in our experiment, the phonon energy is mainly around $2\ \mathrm{meV}$ or $16\ \mathrm{cm^{-1}}$, as shown in Fig.~\ref{f2}(d). The small phonon energy corresponds to the octahedra-twist vibration which is a TO phonon, and the polarons with LO phonons are not observed in the spectra. To investigate the interaction between LXs and different phonons, we measured the temperature-dependent PL spectra, as shown in Fig.~\ref{f4}(a). A large energy shift can be observed with increasing temperature. This again excludes that multiple peaks are due to the FP modes, because the refractive index which determines the FP modes usually changes slowly at low temperature \cite{doi:10.1021/acs.jpclett.7b01740}. The rapid decay of peak intensity shows the decrease of the binding capacity of defects. The intensity ratio between the polaron and ZPL increases with increasing temperature (Fig.~\ref{f4}(b)). This is due to that phonon assisted excitonic transition increases with exciton kinetic energy \cite{HOPFIELD1959110,PhysRev.171.935,Zhang_2001}. Due to Hopfield's theory \cite{HOPFIELD1959110} the intensity ratio is proportional to $\sqrt{T}$ where $T$ is the temperature. Our result agrees well with the theory as shown with the fitted result (gray line in Fig.~\ref{f4}(b)). More importantly, the temperature-dependent spectra provide the way to investigate the interaction to LO phonons which was not directly observed in our case.

The ZPL linewidth mainly originates from the inhomogeneous line broadening and the interaction with the LO phonon \cite{doi:10.1021/acs.nanolett.8b01523}. The linewidth variation with the temperature can be expressed by the Fröhlich formalism
\[\Gamma(T)=\Gamma_{inh}+\frac{\Gamma _{LO}}{e^{E_{LO}/k_{b}T}-1},\]
where $\Gamma_{inh}$ is the inhomogeneous line broadening at 0 K, $\Gamma_{LO}$ is the LO phonon interaction strength to excitons and $E_{LO}$ is the LO phonon energy. The fitted result is shown in Fig.~\ref{f4}(c). The fitted phonon energy $E_{LO}=5.4\ \mathrm{meV}$ ($44\ \mathrm{cm^{-1}}$) corresponds well with the octahedra-distorting vibration (LO/TO phonon) \cite{doi:10.1021/acs.nanolett.7b00064,doi:10.1021/acs.jpclett.7b02979}. The interesting result is the fitted interaction strength $\Gamma_{LO}=13\ \mathrm{meV}$, which is quite small compared to the value over $100\ \mathrm{meV}$ in bulk HOIPs or other materials \cite{Yu2010Fundamentals,PhysRevB.38.13377,C6MH00275G,Wright2016,PhysRevB.58.16333,PhysRevMaterials.2.064605}. Therefore, these results demonstrate that the LXs in HOIPs prefer to interact with TO phonons rather than LO phonons. Generally in semiconductors, the FX, the phonon and the defect could interact with each other and alter optoelectronic properties such as ionization rates, carrier velocities and recombination channels \cite{PhysRevLett.39.716,PhysRevLett.48.1281,TOYOZAWA19837,Maciel2008}. Previous works have proposed that in HOIPs, the polarons with carriers and LO phonons prevent the interaction between carriers and defects \cite{Zhu1409,doi:10.1021/acs.jpclett.5b02462,Miyatae1701217}. Here we demonstrate that defects prefer to interact to TO phonons. This result gives a further explanation to why the polarons with carriers and LO phonons rarely interact with defects. Overall, the FXs and LXs have different vibration modes (different phonons). Thus, the phonon-assisted interaction between defects and FXs or carriers should be much smaller than that in other materials. This significant difference explains the high tolerance of defects in HOIPs, as an important feature related to various attractive optoelectronic properties.
%involves

The interaction between LXs and TO phonons might be ascribed to the unique orthorhombic structure of HOIPs. As shown in Fig.~\ref{f1}(a), the MAs fill the voids between the halogen octahedras. The ionic bond between the MA and the halogen octahedras is relatively weak compared to the covalent bond between halogen atoms and Pb atoms. Meanwhile, the MA has a small size and can even rotate in the voids \cite{doi:10.1021/acs.nanolett.6b01218}. Thus, many MAs are absent during the synthesization, and the energy levels in MA-absent defects also correspond well with the energy of LXs in our work \cite{Chen_2018}. The halogen atoms around the defect will prefer to vibrate along the direction towards the absent MA, resulting in the octahedra-twist vibrations which are TO phonon modes. Therefore, the LXs prefer to interact with TO phonons due to the unique lattice structure of HOIPs.

\section{Conclusion}

In conclusion, the LXs are identified in $\mathrm{CH_{3}NH_{3}PbBr_{3}}$ nanowires with very narrow linewidth down to 226 $\mu$eV. Based on the narrow linewidth, multiple hot polarons originating from the interaction between LXs and TO phonons have been successfully observed , which has been confirmed with the excitation-power-dependent PL, magneto-PL and temperature-dependent PL. Furthermore, the LXs were demonstrated prefer to interact with TO phonons rather than LO phonons. This result indicates that the defects in HOIPs have different coupled vibration modes compared to FXs. This difference gives a physical explanation of high tolerance of defects in HOIPs from the point of phonons. Additionally, the LXs with narrow linewidth is also a potential quantum emitter \cite{doi:10.1063/1.1803624,PhysRevMaterials.3.051001,Dolde2013,Srivastava2015}, thus, more applications can be expected in quantum information processing based on perovskites in the future.

\section{Conflict of Interest}
The authors declare no conflict of interest.

% Give an acknowledgement if you have to thank someone, are
% supported by someone or something like that:
\section{Acknowledgments}
This work was supported by the National Natural Science Foundation of China under Grants No. 11934019, No. 11721404, No. 51761145104, No. 61675228, and No, 11874419, the Strategic Priority Research Program (Grant No. XDB28000000), the Instrument Developing Project (Grant No. YJKYYQ20180036), the Interdisciplinary Innovation Team of the Chinese Academy of Sciences, and the Key RD Program of Guangdong Province (Grant No.2018B030329001).

% The authors' biographies are given inside an \authorbox command
% each, all together enclosed in this environment.  Here is the
% syntax:
% \authorbox{<name>}{Authorname}{Text}
% where "<name>" is the Filename of the portrait picture, if any; it
% should consist of the author's last name prefixed with "cv".
% "Authorname" is this respective author's name being bolded as the
% first word(s) of the biography text, the rest of which follows in
% the third argument.
%\begin{biographies}
%  \authorbox{cv}{}{}
%  \authorbox{cv}{}{}
%  \authorbox{cv}{}{}
%\end{biographies}
%
% Give your bibliography below.  If you are using BibTeX please make
% sure to enclose all .bib files needed!
\providecommand{\WileyBibTextsc}{}
\let\textsc\WileyBibTextsc
\providecommand{\othercit}{}
\providecommand{\jr}[1]{#1}
\providecommand{\etal}{~et~al.}


\begin{thebibliography}{[10]}

\bibitem{doi:10.1002/adma.201502294}% article
 \textsc{J.~Berry},  \textsc{T.~Buonassisi},  \textsc{D.\,A. Egger},
  \textsc{G.~Hodes},  \textsc{L.~Kronik},  \textsc{Y.\,L. Loo},
  \textsc{I.~Lubomirsky},  \textsc{S.\,R. Marder},  \textsc{Y.~Mastai},
  \textsc{J.\,S. Miller},  \textsc{D.\,B. Mitzi},  \textsc{Y.~Paz},
  \textsc{A.\,M. Rappe},  \textsc{I.~Riess},  \textsc{B.~Rybtchinski},
  \textsc{O.~Stafsudd},  \textsc{V.~Stevanovic},  \textsc{M.\,F. Toney},
  \textsc{D.~Zitoun},  \textsc{A.~Kahn},  \textsc{D.~Ginley},  and
  \textsc{D.~Cahen},
 \jr{Adv. Mater.} \textbf{27}(35), 5102--5112 (2015).


\bibitem{Brenner2016}% article
 \textsc{T.\,M. Brenner},  \textsc{D.\,A. Egger},  \textsc{L.~Kronik},
  \textsc{G.~Hodes},  and  \textsc{D.~Cahen},
 \jr{Nat. Rev. Mater.} \textbf{1}, 15007 (2016).


\bibitem{Li2017}% article
 \textsc{W.~Li},  \textsc{Z.~Wang},  \textsc{F.~Deschler},  \textsc{S.~Gao},
  \textsc{R.\,H. Friend},  and  \textsc{A.\,K. Cheetham},
 \jr{Nat. Rev. Mater.} \textbf{2}, 16099 (2017).


\bibitem{Heo2013}% article
 \textsc{J.\,H. Heo},  \textsc{S.\,H. Im},  \textsc{J.\,H. Noh},
  \textsc{T.\,N. Mandal},  \textsc{C.\,S. Lim},  \textsc{J.\,A. Chang},
  \textsc{Y.\,H. Lee},  \textsc{H.\,j. Kim},  \textsc{A.~Sarkar},
  \textsc{M.\,K. Nazeeruddin},  \textsc{M.~Gr{\"a}tzel},  and  \textsc{S.\,I.
  Seok},
 \jr{Nat. Photonics} \textbf{7}, 486--491 (2013).


\bibitem{Jeon2014}% article
 \textsc{N.\,J. Jeon},  \textsc{J.\,H. Noh},  \textsc{Y.\,C. Kim},
  \textsc{W.\,S. Yang},  \textsc{S.~Ryu},  and  \textsc{S.\,I. Seok},
 \jr{Nat. Mater.} \textbf{13}, 897--903 (2014).


\bibitem{Xing2014}% article
 \textsc{G.~Xing},  \textsc{N.~Mathews},  \textsc{S.\,S. Lim},
  \textsc{N.~Yantara},  \textsc{X.~Liu},  \textsc{D.~Sabba},
  \textsc{M.~Gr{\"a}tzel},  \textsc{S.~Mhaisalkar},  and  \textsc{T.\,C. Sum},
 \jr{Nat. Mater.} \textbf{13}, 476--480 (2014).


\bibitem{C4CS00458B}% article
 \textsc{Y.~Zhao} and  \textsc{K.~Zhu},
 \jr{Chem. Soc. Rev.} \textbf{45}, 655--689 (2016).


\bibitem{doi:10.1021/acs.jpclett.9b00658}% article
 \textsc{L.~Wang},  \textsc{L.~Meng},  \textsc{L.~Chen},  \textsc{S.~Huang},
  \textsc{X.~Wu},  \textsc{G.~Dai},  \textsc{L.~Deng},  \textsc{J.~Han},
  \textsc{B.~Zou},  \textsc{C.~Zhang},  and  \textsc{H.~Zhong},
 \jr{J. Phys. Chem. Lett.} \textbf{10}(0), 3248--3253 (2019).


\bibitem{doi:10.1021/acs.nanolett.6b02688}% article
 \textsc{D.\,N. Dirin},  \textsc{L.~Protesescu},  \textsc{D.~Trummer},
  \textsc{I.\,V. Kochetygov},  \textsc{S.~Yakunin},  \textsc{F.~Krumeich},
  \textsc{N.\,P. Stadie},  and  \textsc{M.\,V. Kovalenko},
 \jr{Nano Lett.} \textbf{16}(9), 5866--5874 (2016).


\bibitem{doi:10.1021/acsenergylett.7b00547}% article
 \textsc{H.~Huang},  \textsc{M.\,I. Bodnarchuk},  \textsc{S.\,V. Kershaw},
  \textsc{M.\,V. Kovalenko},  and  \textsc{A.\,L. Rogach},
 \jr{ACS Energy Lett.} \textbf{2}(9), 2071--2083 (2017).


\bibitem{PhysRevB.97.134412}% article
 \textsc{C.~Zhang},  \textsc{D.~Sun},  \textsc{Z.\,G. Yu},  \textsc{C.\,X.
  Sheng},  \textsc{S.~McGill},  \textsc{D.~Semenov},  and  \textsc{Z.\,V.
  Vardeny},
 \jr{Phys. Rev. B} \textbf{97}, 134412 (2018).


\bibitem{doi:10.1021/acs.jpclett.7b02979}% article
 \textsc{O.\,A. Lozhkina},  \textsc{V.\,I. Yudin},  \textsc{A.\,A. Murashkina},
   \textsc{V.\,V. Shilovskikh},  \textsc{V.\,G. Davydov},
  \textsc{R.~Kevorkyants},  \textsc{A.\,V. Emeline},  \textsc{Y.\,V.
  Kapitonov},  and  \textsc{D.\,W. Bahnemann},
 \jr{J. Phys. Chem. Lett.} \textbf{9}(2), 302--305 (2018).


\bibitem{doi:10.1146/annurev-physchem-040215-112222}% article
 \textsc{L.\,M. Herz},
 \jr{Annu. Rev. Phys. Chem.} \textbf{67}(1), 65--89 (2016).


\bibitem{RevModPhys.89.015003}% article
 \textsc{F.~Giustino},
 \jr{Rev. Mod. Phys.} \textbf{89}, 015003 (2017).


\bibitem{doi:10.1021/acsnano.7b05033}% article
 \textsc{C.\,M. Iaru},  \textsc{J.\,J. Geuchies},  \textsc{P.\,M. Koenraad},
  \textsc{D.~Vanmaekelbergh},  and  \textsc{A.\,Y. Silov},
 \jr{ACS Nano} \textbf{11}(11), 11024--11030 (2017).


\bibitem{Zhu1409}% article
 \textsc{H.~Zhu},  \textsc{K.~Miyata},  \textsc{Y.~Fu},  \textsc{J.~Wang},
  \textsc{P.\,P. Joshi},  \textsc{D.~Niesner},  \textsc{K.\,W. Williams},
  \textsc{S.~Jin},  and  \textsc{X.\,Y. Zhu},
 \jr{Science} \textbf{353}(6306), 1409--1413 (2016).


\bibitem{doi:10.1021/acs.jpclett.5b02462}% article
 \textsc{X.\,Y. Zhu} and  \textsc{V.~Podzorov},
 \jr{J. Phys. Chem. Lett.} \textbf{6}(23), 4758--4761 (2015).


\bibitem{Miyatae1701217}% article
 \textsc{K.~Miyata},  \textsc{D.~Meggiolaro},  \textsc{M.\,T. Trinh},
  \textsc{P.\,P. Joshi},  \textsc{E.~Mosconi},  \textsc{S.\,C. Jones},
  \textsc{F.~De~Angelis},  and  \textsc{X.\,Y. Zhu},
 \jr{Sci. Adv.} \textbf{3}(8) (2017).


\bibitem{C6CP03474H}% article
 \textsc{A.\,M.\,A. Leguy},  \textsc{A.\,R. Goñi},  \textsc{J.\,M. Frost},
  \textsc{J.~Skelton},  \textsc{F.~Brivio},  \textsc{X.~Rodríguez-Martínez},
  \textsc{O.\,J. Weber},  \textsc{A.~Pallipurath},  \textsc{M.\,I. Alonso},
  \textsc{M.~Campoy-Quiles},  \textsc{M.\,T. Weller},  \textsc{J.~Nelson},
  \textsc{A.~Walsh},  and  \textsc{P.\,R.\,F. Barnes},
 \jr{Phys. Chem. Chem. Phys.} \textbf{18}, 27051--27066 (2016).


\bibitem{doi:10.1021/acs.nanolett.7b00064}% article
 \textsc{M.~Fu},  \textsc{P.~Tamarat},  \textsc{H.~Huang},  \textsc{J.~Even},
  \textsc{A.\,L. Rogach},  and  \textsc{B.~Lounis},
 \jr{Nano Lett.} \textbf{17}(5), 2895--2901 (2017).


\bibitem{doi:10.1021/acsnano.5b01154}% article
 \textsc{F.~Zhang},  \textsc{H.~Zhong},  \textsc{C.~Chen},  \textsc{X.\,g. Wu},
   \textsc{X.~Hu},  \textsc{H.~Huang},  \textsc{J.~Han},  \textsc{B.~Zou},  and
   \textsc{Y.~Dong},
 \jr{ACS Nano} \textbf{9}(4), 4533--4542 (2015).


\bibitem{doi:10.1002/cnma.201700034}% article
 \textsc{F.~Zhang},  \textsc{C.~Chen},  \textsc{S.\,V. Kershaw},
  \textsc{C.~Xiao},  \textsc{J.~Han},  \textsc{B.~Zou},  \textsc{X.~Wu},
  \textsc{S.~Chang},  \textsc{Y.~Dong},  \textsc{A.\,L. Rogach},  and
  \textsc{H.~Zhong},
 \jr{ChemNanoMat} \textbf{3}(5), 303--310 (2017).


\bibitem{Cho2013}% article
 \textsc{C.\,H. Cho},  \textsc{C.\,O. Aspetti},  \textsc{J.~Park},  and
  \textsc{R.~Agarwal},
 \jr{Nat. Photonics} \textbf{7}, 285--289 (2013).


\bibitem{doi:10.1021/acs.nanolett.5b00109}% article
 \textsc{H.~Kawai},  \textsc{G.~Giorgi},  \textsc{A.~Marini},  and
  \textsc{K.~Yamashita},
 \jr{Nano Lett.} \textbf{15}(5), 3103--3108 (2015).


\bibitem{doi:10.1021/jacs.6b08175}% article
 \textsc{D.\,B. Straus},  \textsc{S.~Hurtado~Parra},  \textsc{N.~Iotov},
  \textsc{J.~Gebhardt},  \textsc{A.\,M. Rappe},  \textsc{J.\,E. Subotnik},
  \textsc{J.\,M. Kikkawa},  and  \textsc{C.\,R. Kagan},
 \jr{J. Am. Chem. Soc.} \textbf{138}(42), 13798--13801 (2016).


\bibitem{doi:10.1021/acsenergylett.7b00862}% article
 \textsc{J.\,M. Frost},  \textsc{L.\,D. Whalley},  and  \textsc{A.~Walsh},
 \jr{ACS Energy Lett.} \textbf{2}(12), 2647--2652 (2017).


\bibitem{doi:10.1021/acs.jpclett.6b02046}% article
 \textsc{K.~Zheng},  \textsc{M.~Abdellah},  \textsc{Q.~Zhu},  \textsc{Q.~Kong},
   \textsc{G.~Jennings},  \textsc{C.\,A. Kurtz},  \textsc{M.\,E. Messing},
  \textsc{Y.~Niu},  \textsc{D.\,J. Gosztola},  \textsc{M.\,J. Al-Marri},
  \textsc{X.~Zhang},  \textsc{T.~Pullerits},  and  \textsc{S.\,E. Canton},
 \jr{J. Phys. Chem. Lett.} \textbf{7}(22), 4535--4539 (2016).


\bibitem{doi:10.1002/aenm.201602174}% article
 \textsc{D.~Raiser},  \textsc{S.~Mildner},  \textsc{B.~Ifland},
  \textsc{M.~Sotoudeh},  \textsc{P.~Blöchl},  \textsc{S.~Techert},  and
  \textsc{C.~Jooss},
 \jr{Adv. Energy Mater.} \textbf{7}(12), 1602174 (2017).


\bibitem{C8EE03369B}% article
 \textsc{F.~Zheng} and  \textsc{L.\,W. Wang},
 \jr{Energy Environ. Sci.} \textbf{12}, 1219--1230 (2019).


\bibitem{Kazimierczuk2014}% article
 \textsc{T.~Kazimierczuk},  \textsc{D.~Fr{\"o}hlich},  \textsc{S.~Scheel},
  \textsc{H.~Stolz},  and  \textsc{M.~Bayer},
 \jr{Nature} \textbf{514}, 343--347 (2014).


\bibitem{Zhu2015}% article
 \textsc{H.~Zhu},  \textsc{Y.~Fu},  \textsc{F.~Meng},  \textsc{X.~Wu},
  \textsc{Z.~Gong},  \textsc{Q.~Ding},  \textsc{M.\,V. Gustafsson},
  \textsc{M.\,T. Trinh},  \textsc{S.~Jin},  and  \textsc{X.\,Y. Zhu},
 \jr{Nat. Mater.} \textbf{14}, 636--642 (2015).


\bibitem{doi:10.1021/nl9017012}% article
 \textsc{H.\,Y. Li},  \textsc{S.~Rühle},  \textsc{R.~Khedoe},  \textsc{A.\,F.
  Koenderink},  and  \textsc{D.~Vanmaekelbergh},
 \jr{Nano Lett.} \textbf{9}(10), 3515--3520 (2009).


\bibitem{RevModPhys.82.427}% article
 \textsc{R.~Rurali},
 \jr{Rev. Mod. Phys.} \textbf{82}, 427--449 (2010).


\bibitem{PhysRevB.94.125139}% article
 \textsc{U.\,G. Jong},  \textsc{C.\,J. Yu},  \textsc{J.\,S. Ri},
  \textsc{N.\,H. Kim},  and  \textsc{G.\,C. Ri},
 \jr{Phys. Rev. B} \textbf{94}, 125139 (2016).


\bibitem{doi:10.1021/acsnano.6b02734}% article
 \textsc{J.~Tilchin},  \textsc{D.\,N. Dirin},  \textsc{G.\,I. Maikov},
  \textsc{A.~Sashchiuk},  \textsc{M.\,V. Kovalenko},  and
  \textsc{E.~Lifshitz},
 \jr{ACS Nano} \textbf{10}(6), 6363--6371 (2016).


\bibitem{Chen2016}% article
 \textsc{Y.\,C. Chen},  \textsc{P.\,S. Salter},  \textsc{S.~Knauer},
  \textsc{L.~Weng},  \textsc{A.\,C. Frangeskou},  \textsc{C.\,J. Stephen},
  \textsc{S.\,N. Ishmael},  \textsc{P.\,R. Dolan},  \textsc{S.~Johnson},
  \textsc{B.\,L. Green},  \textsc{G.\,W. Morley},  \textsc{M.\,E. Newton},
  \textsc{J.\,G. Rarity},  \textsc{M.\,J. Booth},  and  \textsc{J.\,M. Smith},
 \jr{Nat. Photonics} \textbf{11}, 77--80 (2016).


\bibitem{Cao2016}% article
 \textsc{S.~Cao},  \textsc{J.~Tang},  \textsc{Y.~Sun},  \textsc{K.~Peng},
  \textsc{Y.~Gao},  \textsc{Y.~Zhao},  \textsc{C.~Qian},  \textsc{S.~Sun},
  \textsc{H.~Ali},  \textsc{Y.~Shao},  \textsc{S.~Wu},  \textsc{F.~Song},
  \textsc{D.\,A. Williams},  \textsc{W.~Sheng},  \textsc{K.~Jin},  and
  \textsc{X.~Xu},
 \jr{Nano Res.} \textbf{9}(2), 306--316 (2016).


\bibitem{PhysRevLett.122.087401}% article
 \textsc{C.~Qian},  \textsc{X.~Xie},  \textsc{J.~Yang},  \textsc{K.~Peng},
  \textsc{S.~Wu},  \textsc{F.~Song},  \textsc{S.~Sun},  \textsc{J.~Dang},
  \textsc{Y.~Yu},  \textsc{M.\,J. Steer},  \textsc{I.\,G. Thayne},
  \textsc{K.~Jin},  \textsc{C.~Gu},  and  \textsc{X.~Xu},
 \jr{Phys. Rev. Lett.} \textbf{122}, 087401 (2019).


\bibitem{Skolnick_1983}% article
 \textsc{M.\,S. Skolnick},  \textsc{P.\,J. Dean},  \textsc{M.\,J. Kane},
  \textsc{C.~Uihlein},  \textsc{D.\,J. Robbins},  \textsc{W.~Hayes},
  \textsc{B.~Cockayne},  and  \textsc{W.\,R. MacEwan},
 \jr{J. Phys. C: Solid State Phys.} \textbf{16}(21), L767--L775 (1983).


\bibitem{Zhang_2014}% article
 \textsc{P.~Zhang},  \textsc{F.~Su},  \textsc{J.~Dai},  \textsc{K.~Zhang},
  \textsc{C.~Zhang},  \textsc{W.~Zhou},  \textsc{Q.~Liu},  and  \textsc{L.~Pi},
 \jr{EPL} \textbf{106}(6), 67005 (2014).


\othercit
\bibitem{Yu2010Fundamentals}% book
 \textsc{P.\,Y. Yu} and  \textsc{M.~Cardona},
Fundamentals of semiconductors : physics and materials properties (Springer,
  Heidelberg, Berlin, 2010).


\bibitem{PhysRevB.38.13377}% article
 \textsc{K.~Huang} and  \textsc{B.~Zhu},
 \jr{Phys. Rev. B} \textbf{38}, 13377--13386 (1988).


\bibitem{Wright2016}% article
 \textsc{A.\,D. Wright},  \textsc{C.~Verdi},  \textsc{R.\,L. Milot},
  \textsc{G.\,E. Eperon},  \textsc{M.\,A. P{\'e}rez-Osorio},  \textsc{H.\,J.
  Snaith},  \textsc{F.~Giustino},  \textsc{M.\,B. Johnston},  and
  \textsc{L.\,M. Herz},
 \jr{Nat. Commun.} \textbf{7}, 11755 (2016).


\bibitem{doi:10.1021/acs.jpclett.7b01740}% article
 \textsc{S.~Govinda},  \textsc{B.\,P. Kore},  \textsc{M.~Bokdam},
  \textsc{P.~Mahale},  \textsc{A.~Kumar},  \textsc{S.~Pal},
  \textsc{B.~Bhattacharyya},  \textsc{J.~Lahnsteiner},  \textsc{G.~Kresse},
  \textsc{C.~Franchini},  \textsc{A.~Pandey},  and  \textsc{D.\,D. Sarma},
 \jr{J. Phys. Chem. Lett.} \textbf{8}(17), 4113--4121 (2017).


\bibitem{HOPFIELD1959110}% article
 \textsc{J.~Hopfield},
 \jr{J. Phys. Chem. Solids} \textbf{10}(2), 110 -- 119 (1959).


\bibitem{PhysRev.171.935}% article
 \textsc{B.~Segall} and  \textsc{G.\,D. Mahan},
 \jr{Phys. Rev.} \textbf{171}, 935--948 (1968).


\bibitem{Zhang_2001}% article
 \textsc{X.\,B. Zhang},  \textsc{T.~Taliercio},  \textsc{S.~Kolliakos},  and
  \textsc{P.~Lefebvre},
 \jr{J. Phys.: Condens. Matter} \textbf{13}(32), 7053--7074 (2001).


\bibitem{doi:10.1021/acs.nanolett.8b01523}% article
 \textsc{O.~Pfingsten},  \textsc{J.~Klein},  \textsc{L.~Protesescu},
  \textsc{M.\,I. Bodnarchuk},  \textsc{M.\,V. Kovalenko},  and
  \textsc{G.~Bacher},
 \jr{Nano Lett.} \textbf{18}(7), 4440--4446 (2018).


\bibitem{C6MH00275G}% article
 \textsc{M.~Sendner},  \textsc{P.\,K. Nayak},  \textsc{D.\,A. Egger},
  \textsc{S.~Beck},  \textsc{C.~Müller},  \textsc{B.~Epding},
  \textsc{W.~Kowalsky},  \textsc{L.~Kronik},  \textsc{H.\,J. Snaith},
  \textsc{A.~Pucci},  and  \textsc{R.~Lovrinčić},
 \jr{Mater. Horiz.} \textbf{3}, 613--620 (2016).


\bibitem{PhysRevB.58.16333}% article
 \textsc{A.\,K. Viswanath},  \textsc{J.\,I. Lee},  \textsc{D.~Kim},
  \textsc{C.\,R. Lee},  and  \textsc{J.\,Y. Leem},
 \jr{Phys. Rev. B} \textbf{58}, 16333--16339 (1998).


\bibitem{PhysRevMaterials.2.064605}% article
 \textsc{S.~Neutzner},  \textsc{F.~Thouin},  \textsc{D.~Cortecchia},
  \textsc{A.~Petrozza},  \textsc{C.~Silva},  and  \textsc{A.\,R.
  Srimath~Kandada},
 \jr{Phys. Rev. Materials} \textbf{2}, 064605 (2018).


\bibitem{PhysRevLett.39.716}% article
 \textsc{L.\,R. Testardi},  \textsc{J.\,M. Poate},  \textsc{W.~Weber},
  \textsc{W.\,M. Augustyniak},  and  \textsc{J.\,H. Barrett},
 \jr{Phys. Rev. Lett.} \textbf{39}, 716--719 (1977).


\bibitem{PhysRevLett.48.1281}% article
 \textsc{S.~Makram-Ebeid} and  \textsc{M.~Lannoo},
 \jr{Phys. Rev. Lett.} \textbf{48}, 1281--1284 (1982).


\bibitem{TOYOZAWA19837}% article
 \textsc{Y.~Toyozawa},
 \jr{Physica B+C} \textbf{116}(1), 7 -- 17 (1983).


\bibitem{Maciel2008}% article
 \textsc{I.\,O. Maciel},  \textsc{N.~Anderson},  \textsc{M.\,A. Pimenta},
  \textsc{A.~Hartschuh},  \textsc{H.~Qian},  \textsc{M.~Terrones},
  \textsc{H.~Terrones},  \textsc{J.~Campos-Delgado},  \textsc{A.\,M. Rao},
  \textsc{L.~Novotny},  and  \textsc{A.~Jorio},
 \jr{Nat. Mater.} \textbf{7}, 878--883 (2008).


\bibitem{doi:10.1021/acs.nanolett.6b01218}% article
 \textsc{A.\,J. Neukirch},  \textsc{W.~Nie},  \textsc{J.\,C. Blancon},
  \textsc{K.~Appavoo},  \textsc{H.~Tsai},  \textsc{M.\,Y. Sfeir},
  \textsc{C.~Katan},  \textsc{L.~Pedesseau},  \textsc{J.~Even},  \textsc{J.\,J.
  Crochet},  \textsc{G.~Gupta},  \textsc{A.\,D. Mohite},  and
  \textsc{S.~Tretiak},
 \jr{Nano Lett.} \textbf{16}(6), 3809--3816 (2016).


\bibitem{Chen_2018}% article
 \textsc{C.~Chen},  \textsc{X.~Hu},  \textsc{W.~Lu},  \textsc{S.~Chang},
  \textsc{L.~Shi},  \textsc{L.~Li},  \textsc{H.~Zhong},  and  \textsc{J.\,B.
  Han},
 \jr{J. Phys. D: Appl. Phys.} \textbf{51}(4), 045105 (2018).


\bibitem{doi:10.1063/1.1803624}% article
 \textsc{X.~Xu},  \textsc{D.\,A. Williams},  and  \textsc{J.\,R.\,A. Cleaver},
 \jr{Appl. Phys. Lett.} \textbf{85}(15), 3238--3240 (2004).


\bibitem{PhysRevMaterials.3.051001}% article
 \textsc{Y.~Yu},  \textsc{J.~Dang},  \textsc{C.~Qian},  \textsc{S.~Sun},
  \textsc{K.~Peng},  \textsc{X.~Xie},  \textsc{S.~Wu},  \textsc{F.~Song},
  \textsc{J.~Yang},  \textsc{S.~Xiao},  \textsc{L.~Yang},  \textsc{Y.~Wang},
  \textsc{X.~Shan},  \textsc{M.\,A. Rafiq},  \textsc{B.\,B. Li},  and
  \textsc{X.~Xu},
 \jr{Phys. Rev. Materials} \textbf{3}, 051001 (2019).


\bibitem{Dolde2013}% article
 \textsc{F.~Dolde},  \textsc{I.~Jakobi},  \textsc{B.~Naydenov},
  \textsc{N.~Zhao},  \textsc{S.~Pezzagna},  \textsc{C.~Trautmann},
  \textsc{J.~Meijer},  \textsc{P.~Neumann},  \textsc{F.~Jelezko},  and
  \textsc{J.~Wrachtrup},
 \jr{Nat. Phys.} \textbf{9}, 139--143 (2013).


\bibitem{Srivastava2015}% article
 \textsc{A.~Srivastava},  \textsc{M.~Sidler},  \textsc{A.\,V. Allain},
  \textsc{D.\,S. Lembke},  \textsc{A.~Kis},  and  \textsc{A.~Imamoglu},
 \jr{Nat. Nanotechnol.} \textbf{10}, 491--496 (2015).


\end{thebibliography}
\end{document}